\documentstyle[a4,12pt]{article}
\begin{document}

\vspace*{-1cm}

accepted for publication in {\sl Phys.~Rev.~B}, August 25, 1999\\

\begin{center}
{\Large Influence of the cooperative Jahn-Teller effect\\[0.5ex]
on the transport- and magnetic properties of\\[1ex]
La$_{7/8}$Sr$_{1/8}$MnO$_3$ single crystals}\\[3ex]

P.~Wagner, I.~Gordon, S.~Mangin$^\ddagger$, V.~V.~Moshchalkov,
and Y.~Bruynseraede\\[0.5ex]
Laboratorium voor Vaste-Stoffysica en Magnetisme, Katholieke Universiteit
Leuven, Celestijnenlaan 200 D, 3001 Leuven, Belgium\\[2ex]

L.~Pinsard and A.~Revcolevschi\\[0.5ex]
Laboratoire de Chimie des Solides, Universit\'{e} Paris-Sud,
91405 Orsay C\'{e}dex, France\\[4ex]

{\bf Abstract}\\
\end{center}

The low-doped magnetic perovskite La$_{7/8}$Sr$_{1/8}$MnO$_3$ undergoes
within the paramag\-netic-semi\-con\-duct\-ing phase a first-order
structural transition due to antiferrodistorsive ordering of Jahn-Teller
deformed MnO$_6$ octahedra. This allows to study not only the influence of
the spin configuration on the magneto-transport properties (CMR effect)
but also the role of orbital order and disorder. The orbital ordering
transition (at 269 K in zero magnetic field) causes a doubling of the
resistivity (regardless of the CMR effect in applied magnetic fields) and a
drop of the paramagnetic susceptibility. The latter might be interpreted
in terms of a shrinking of spin polarons. External magnetic fields shift
the ordering transition to lower temperatures according to the field-induced
decrease of the carrier localization. The magnetic field - temperature
phase boundary line was investigated by means of magnetoresistance (up
to 12~T) and pulsed-fields magnetization measurements up to 50~T. The ~
pronounced magnetization anomalies, associated with the phase transition,
vanish for fields exceeding 20~T. This behaviour has been attributed to a
field-induced crossover from antiferrodistorsive order to a
nondistorsive/ferromagnetic orbital configuration.\\[4ex]

\begin{tabular}{lll}
{\bf PACS:} & {\bf 61.50.Ks} & Crystallographic aspects of phase transformations\\
            & {\bf 71.30.+h} & Metal-insulator transitions and other electronic transitions\\
            & {\bf 71.70.Ej} & Spin-orbit coupling, Zeeman and Stark splitting,
              \underline{Jahn-Teller effect}\\
            & {\bf 75.30.Vn} & Colossal magnetoresistance\\[4ex]
\end{tabular}

\begin{tabular}{lll}
Corr.~author: & Dr.~Patrick Wagner  &   \\
   &  Laboratorium voor Vaste -                    &   \\
   &  Stoffysica en Magnetisme                               &   \\
   &  Katholieke Universiteit Leuven   &  Tel. 0032 - 16 - 32 76 46\\
   &  Celestijnenlaan 200 D    &  Fax. 0032 - 16 - 32 79 83\\
   &  B-3001 Leuven / Belgium  &  Patrick.Wagner@fys.kuleuven.ac.be\\[3ex]
\end{tabular}

\vspace*{5mm}

$\ddagger$: present address: Laboratoire de Physique des Mat\'{e}riaux,
Universit\'{e} H.~Poincar\'{e}, \hspace*{5mm} B.~P.~239,
54506 Vandoeuvre-les-Nancy C\'{e}dex, France

\newpage

{\bf 1.~Introduction}\\

Rare-earth (RE) manganites with partial divalent (D) substitution on the
rare-earth site, i.~e.~RE$_{1-x}$D$_x$MnO$_3$, show a transition from a
paramagnetic-semiconducting phase to a ferromagnetic-quasimetallic state.
This insulator-metal transition can be tuned by an external magnetic field,
resulting in a colossal negative magnetoresistance effect (CMR)\cite{Kusters}.
An overview on the physics of manganite materials in general and the relevant
electronic transitions can be found in the review articles by Ramirez
\cite{Ramirez} and by Imada et al.~\cite{Imada}. The relationship between
ferromagnetic spin alignment and enhanced charge carrier mobility between
neighbouring Mn$^{3+}$ and Mn$^{4+}$ ions has been described in terms of
the double-exchange model \cite{deGennes} and its recent extensions, taking
into account the role of electron-phonon coupling and the existence of a
Berry phase \cite{Millis,Mueller}. Also a Mott-type hopping model, in which
the effective barrier depends on the mutual spin orientation at the hopping
sites, gives a correct description of the CMR effect in the para- and in the
ferromagnetic state \cite{Wagner}. The charge carriers are quite close to
the localized state and the shielded Coulomb repulsion between them
results, for special commensurate substitution ratios, e.g.~$x = 1/8, 1/4,
1/2$, in an additional phase transition to a charge-ordered, poorly
conducting state, being commonly described as 'Wigner-' or 'charge-crystal'
\cite{Care}. Prototypes of charge ordering compounds are single crystals of
Nd$_{0.5}$ (Pr$_{0.5}$)Sr$_{0.5}$MnO$_3$ \cite{Tomioka,Kawano97} and
La$_{7/8}$Sr$_{1/8}$MnO$_3$ \cite{Pinsard,Uhlenbruck,Niemoeller}. The
superstructure of ordered charges (Mn ions of different valency) was in
both cases confirmed by neutron- and hard-x-ray diffraction. Fundamentally
different however is the behaviour of these Wigner phases under the
influence of external magnetic fields: the 50\% - doped compounds are
antiferromagnetic insulators at low temperature, and applying magnetic
fields destroys the antiferromagnetic alignment as well as delocalizes the
charge carriers, thus resulting in a negative magnetoresistance of several
orders of magnitude \cite{Tomioka}. At low temperature the
La$_{7/8}$-crystals are ferromagnetic insulators, and the transition to the
ordered state shifts to higher temperatures under external fields, giving
rise to a slightly positive magnetoresistance \cite{Uhlenbruck}.\\
\hspace*{5mm} Another important feature of manganites, besides magnetism
and charge ordering, is the Jahn-Teller (JT) distortion of the
Mn$^{3+}$O$_6^{2-}$ octahedra \cite{Millis}. A review on the Jahn-Teller
effect in general can be found in ref.~\cite{Gehring}. Its most common
appearance is a volume-conserving elongation of the octahedra along one
axis and a compression along the other two axes. This lifts the energetic
degeneracy of the 3d-e$_g$ level and stabilizes the 3d$_{3z^2-r^2}$ orbital
with respect to the 3d$_{x^2-y^2}$ state. In the following we will refer to
these orbitals simply as '$3z^2-r^2$' and '$x^2-y^2$'. At high temperatures
the distortion is not reflected in the variation of the lattice constants
due to a mixed occupation of these orbitals, which can moreover be randomly
oriented along the main crystalline axes. There is also a breathing-type
oscillation between the elongated type and a compressed version of the JT
distortion, which favours the $x^2-y^2$ orbital rather than $3z^2-r^2$.
This disordered and fluctuating state can be described as a 'dynamic
JT-state', abbreviated by 'DJT'. The JT-ordering temperature T$_{JT}$ is
characterized by a preferential occupation of $3z^2-r^2$ orbitals and the
distortion becomes static with a coherent orientation of the elongated
axes throughout the sample. This state is in the following described as
'cooperative JT effect' and abbreviated as 'CJT'. Pure LaMnO$_3$ shows
hereby the so-called antiferrodistorsive orbital order
\cite{Goodenough,Murakami}, which minimizes the increase of elastic
energy on a macroscopic length scale. This peculiar type of order is
preserved in doped manganites with a maximum Mn$^{4+}$ content of $x= 0.15$
\cite{Niemoeller}, at the expense of a lowering of T$_{JT}$, e.~g.~from
790~K for the undoped system to 269~K for $x = 1/8$
\cite{Pinsard,Rodriguez}.\\
\hspace*{5mm} The main objective of this article is to investigate
the relationship between the resistive/magnetic properties of
La$_{7/8}$Sr$_{1/8}$MnO$_3$ single crystals and the orbital configuration of
the $e_g$ electrons. These electrons mediate the transport- as well as the
magnetic properties and the difference between a random configuration and a
static arrangement of electron orbitals should have a significant impact on
charge transfer and magnetic interactions on a macroscopic scale. More
specifically, we found that the freezing of the orbital configuration into
a regular pattern enhances the resistivity by a roughly a factor of two,
accompanied by a simultaneous drop of the paramagnetic susceptibility. Both
observations will be analyzed and explained on grounds of an orbital
ordering model, which is based on the magnetic- and charge-transfer
interactions between the diluted Mn$^{4+}$ ions and the surrounding
nearest neighbours of Mn$^{3+}$.\\

{\bf 2.~Experimental}\\

Single crystals of La$_{7/8}$Sr$_{1/8}$MnO$_3$ were prepared from sintered
polycrystalline rods by the floating zone method, with the growth direction
approximately along the $b$-axis \cite{Anane95}. Due to the relative
smallness of the orthorhombic deviation from a perfectly cubic structure we
can, however, not exclude the possibility of some microtwinning in the
crystals. The sample used for our measurements was cut to the dimensions
$1 \times 2 \times 3$~mm$^3$. Four gold contacts were evaporated onto the
$2 \times 3$~mm$^2$ top side and annealed in air for 60 min at 600$^o$C.
The measuring current was flowing in the Lorentz-force free configuration
parallel to the applied magnetic field. The magnetoresistivity measurements
were performed in a temperature range between 1.5~K and 300~K in a cryostat
equipped with a superconducting magnet coil generating fields up to 12~T
and magnetization was measured in a SQUID magnetometer in fields up to
5~T. For magnetization studies at higher magnetic fields we employed the pulsed
fields setup described in ref.~\cite{Herlach}, allowing to measure
magnetization induced by field pulses up to 50~T on a timescale
of 10 - 20 ms. The detected signal is hereby the voltage induced in highly
sensitive pick-up coils, being proportional to the time derivative of the
magnetization $\partial M / \partial t$, which is electronically integrated
to $M(B)$. The filling factor and temperature-dependent sensitivity were
calibrated according to the absolute magnetization values obtained by the
SQUID. The nominal temperatures of these pulsed-field measurements in the
considered temperature range (between 200~K and 300~K) are accurate
within $\pm$ 4~K. \\

{\bf 3.~Results and Discussion}\\

{\bf 3.1~Correlations between the structure and resistive/magnetic
properties}\\

The temperature dependence of the lattice constants in zero external field
is given in Fig.~1a), together with the temperature-dependent resistivity
(Fig.~1b) and the sample magnetization in Fig.~1c). The structural data are
adopted from the x-ray- and synchrotron radiation studies by Niem\"oller et
al.~\cite{Niemoeller} and agree with neutron diffraction results obtained
on the same compound by Pinsard et al.~\cite{Pinsard} and by Kawano
et al.~\cite{Kawano96}. The structure at room-temperature is pseudocubic
with a slight orthorhombic distortion, which becomes notably stronger in the
antiferrodistorsive state. The $a-$ and $b$-axis are hereby expanded and the
$c$-axis becomes compressed. The expansion of $a$ is much less pronounced
than the $b$-expansion, which is somewhat uncommon. This might be related
to an interplay of the JT distortion with the rotation and tilt of MnO$_6$
octahedra with respect to the La$_{7/8}$Sr$_{1/8}$ lattice, or to additional
lattice distortions induced by the relatively small Mn$^{4+}$ ions. At low
temperatures there is a reentrant structural transition to the very same
lattice parameters found at room temperature, and a possible orbital
configuration, consistent with these structural data, will be suggested
below. The two structural transitions, together with the magnetic transition,
allow to distinguish four different temperature regimes (see Fig.~1):\\

{\bf (i)} At room temperature the system is paramagnetic/semiconducting, and the
the conductivity is usually ascribed to thermally activated polaron hopping.
The JT distortion is not seen in the 'macroscopic' structural data, and is
therefore of the 'disordered' or 'dynamic' type. This is schematically shown
in Fig.~2, where the 3d-$e_g$ electrons of Mn$^{3+}$ occupy $x^2-y^2$-
as well as $3z^2-r^2$ orbitals, oriented along random axes. Moreover, there
is a breathing-type phonon mode, which allows for an oscillation between
these two orbital types for a given Mn$^{3+}$ site. The axes in Fig.~2
(and in Figs.~3,4) are denoted as '$x^\star,y^\star,z^\star$',
to discriminate them from those of the pseudocubic unit cell '$a,b,c$',
and from the 'x,y,z' coordinates, employed for the description of the shape
of individual orbitals.\\

{\bf (ii)} At T = 269~K, in the following denoted as Jahn-Teller temperature
T$_{JT}$, a structural transition arises, which is attributed to an
antiferrodistorsive ordering of JT-elongated MnO$_6$ octahedra
\cite{Pinsard}, i.e.~the $e_g$ electrons occupy predominantly $3z^2-r^2$
orbitals. A graphic representation of this peculiar type of orbital order
is given in Fig.~3. The expanded pseudocubic axes $a$ and $b$ correspond to
the diagonals between $x^\star$ and $y^\star$, while the compressed $c$ axis
is oriented along the $z^\star$ direction. This structure of orbitals is
equivalent to the 'resonant x-ray scattering' results by Murakami
et al.~on undoped LaMnO$_3$ \cite{Murakami}. It is noteworthy that
the mutual orbital orientation at neighbouring Mn sites causes ferromagnetic
correlations within the $x^\star y^\star$-plane, and antiferromagnetic
superexchange in the perpendicular direction \cite{Goodenough}. In LaMnO$_3$
these interactions are weak and lead to A-type antiferromagnetism below
T$_N$ = 140~K, indicating already that the low-temperature orbital structure
of La$_{7/8}$Sr$_{1/8}$MnO$_3$ deviates from the antiferrodistorsive
type. The orbital ordering at 269~K is a first-order phase transition
(compare the specific heat results \cite{Uhlenbruck}) and accompanied
by a sudden increase of resistivity (roughly by a factor of two) and
an instantaneous drop of the paramagnetic susceptibility by 20 \%\ .
These two interrelated effects will be discussed on grounds of the
orbital structure in Sections 3.2 and 3.3. Effects with similar
appearance (enhancement of resistivity and drop of magnetization due to
a structural transition) were also observed within the paramagnetic phase
of La$_{0.83}$Sr$_{0.17}$MnO$_3$ \cite{Neumeier} and in the ferromagnetic
phase of La$_{0.825}$Sr$_{0.175}$Mn$_{0.94}$Mg$_{0.06}$O$_3$ \cite{Anane96}.
The origin of the structural transition might be different
from antiferrodistorsive order due to the relatively high Mn$^{4+}$
content, exceeding the threshhold value of 15~\%\ \cite{Niemoeller}).\\

{\bf (iii)} From the extrapolation of the inverse paramagnetic susceptibility to
zero one finds a ferromagnetic Curie temperature T$_C$ = 188 K. At T$_C$ the
colossal negative magnetoresistance effect is maximum and the resistivity
decreases below this temperature in a quasimetallic manner, meaning that
$d\rho / dT > 0$ while the absolute $\rho$ values are unusually high
compared to common metals. The antiferrodistorsive orthorhombicity starts
to decrease gradually at T$_C$ and disappears around 150 K. This indicates
that there is a competition between ferromagnetism (induced by the little
Mn$^{4+}$ content) and the antiferrodistorsive structure, favouring an
A-type AFM. We note that the magnetization deviates from that of
an ideal ferromagnet because the absolute value stabilizes at 3/4 of the
low-temperature value. This points to a possible phase separation into
ferromagnetic and para-/or antiferromagnetic areas - a phenomenon, which
is observed in a wide variety of manganite compounds \cite{Moreo}.\\

{\bf (iv)} At 147 K the resistivity increases spontaneously by roughly a
factor of two, which is associated with a charge-ordering transition to a
Wigner-crystal state. The charge order was demonstrated by the observation
of superlattice reflections in x-ray diffraction \cite{Niemoeller}. The
charge-ordering transition is also accompanied by a sudden jump in
magnetization to the actual low-temperature value and by the vanishing of
the macroscopically observed JT distortion. The Wigner state is not perfectly
insulating and the resistivity increase between 146 K (right below the
Wigner transition) and 75~K can best be described by Shklovskii-Efros (SE)
hopping with $\rho (T) \propto \exp\{(T_0/T)^{1/2}\}$ \cite{Efros}.
This hopping mechanism corresponds to variable range hopping with
a soft Coulomb gap in the density of states. Other hopping processes, like
thermally activated (polaron) hopping, Mott's variable range hopping,
or cascade hopping, do not apply. From the SE description of the resistivity
in the temperature regime between 75~K and 146~K (see fit function in
Fig.~1b) we could extract $T_0$ = 1.26 $\cdot$ 10$^4$~K, where $T_0$
is given by $k_B T_0 = 2.8 e^2 / (4 \pi \epsilon_0 \epsilon_L L)$, with
$\epsilon_L$ being the dielectric constant of the lattice and $L$ the
carrier localization length \cite{Efros}. The resulting
$L$ = 470 \AA\ / $\epsilon_L$ seems of the correct magnitude, since the
dielectric constant of the ionic perovskites easily achieves values in the
range of $10^2$, e.g.~up to 300 for the isostructural SrTiO$_3$. More
precise data on $\epsilon_L$ of manganites are to our knowledge not yet
available. The resistivity increase below 75~K is weaker than described
by any of the afore-mentioned hopping processes, including SE hopping. The
finite conductivity for T $\rightarrow$ 0 means that the Wigner crystal
exhibits imperfections in form of non-localized charge carriers, which
might result from slight deviations from the exact 1/8 doping ratio.
Interestingly enough, this remanent conductivity is also found in the
charge-ordered Pr- and Nd-manganites with 50 \%\ strontium doping
\cite{Tomioka,Kawano97}. The resistivity increase below 40~K (Fig.~1b)
scales with $- \log T$, pointing either to a Kondo-type problem or to a
resistivity contribution caused by electron-electron
interactions \cite{Aliev}.\\
      \hspace*{5mm} In the low-temperature limit we found a magnetic moment
per Mn ion of 3.75 $\mu_B$ (at 10~K, 5~T). This is very close to the
spin-only value for a mixture of Mn$_{7/8}^{3+}$ ($J = s = 2$) and
Mn$_{1/8}^{4+}$ ($J = s = 3/2$) and a gyromagnetic ratio of $g = 2$,
resulting in 3.88 $\mu_B$. This is, within the precision of the measurement,
compatible with ferromagnetic spin alignment, although neutron
diffraction gave possible evidence for a small spin canting
\cite{Pinsard}.\\

Both observations - ferromagnetism and recovery of the DJT-lattice
parameters - suggest that the orbital structure at low temperatures is
different from the antiferrodistorsive configuration. The possible guess
that lowering temperature transfers the static order back to a
dynamic/disordered JT-state, as at high temperatures, is clearly
contra-intuitive. Furthermore, XAFS-studies on La$_{2/3}$Ca$_{1/3}$MnO$_3$
have shown that there are two peaks in the distribution function of Mn-O
bond lengths for the paramagnetic state (attributed to the JT distortion),
which merge to a single peak for the ferromagnetic quasimetal \cite{Tyson}.
This is interpreted in the sense that the $e_g$  electrons become
delocalized and need therefore not to profit from the formation of JT
distortions on the scale of individual unit cells. In the case of
La$_{7/8}$Sr$_{1/8}$MnO$_3$ at low temperatures, the electrons are, despite
of ferromagnetism, well localized and the JT-distortion is nevertheless not
seen in the XRD data of Fig.~1a. This may be interpreted in terms of the
orbital structure sketched in Fig.~4, which agrees with a 'G-type orbital
order', in the scheme of Maezono et.~al.~\cite{Maezono}. Firstly, this
orbital configuration results in a ferromagnetic nearest neighbour coupling
between Mn$^{3+}$ ions for all directions \cite{Goodenough}, while the
coupling between Mn$^{3+}$ and the low-concentrated Mn$^{4+}$ is {\sl per se}
mainly ferromagnetic. Secondly, this structure provides an exact
compensation of the elongated and compressed axes of the JT-distorted
octahedra on a macroscopic scale: the $3z^2-r^2$ orbitals (50 \%\ occupation)
are surrounded by octahedra, which are elongated by a value
$+ \delta$ along the orbital's axis. Since the JT-effect is in first
approximation volume-conserving, the contraction along the two perpendicular
axes has to be $\approx - \delta/2$. The corresponding figures for the
$x^2-y^2$ orbitals (50 \%\ occupation) are an expansion by $+ \delta/2$
within the plane of these orbitals, and a compression by $- \delta$ in the
perpendicular direction. These distortions compensate precisely, in the
structure of Fig.~4, along the three spatial directions, provided that we
can ignore local distortions caused by the low-concentrated Mn$^{4+}$.\\

{\bf 3.2~Influence of the CJT transition on the resistivity}\\

We performed $\rho(B)$ measurements at constant temperatures around
T$_{JT}$, see Fig.~5. Besides the usual CMR behaviour, there is a sharp
negative magnetoresistance effect upon crossing the boundary of the
antiferrodistorsive phase, which is stable in low fields, and entering the
disordered DJT phase in higher magnetic fields. The strong hysteresis effect (being
especially pronounced at 268~K) gives further confirmation of the
first-order nature of the CJT transition. Upon lowering the temperature
the transition shifts to higher fields with a shrinking of the
hysteresis width. The shape of the phase-boundary line will be
discussed in Sect.~3.4. Almost independent of the temperature is,
however, the relative height of the resisitive jump with
2.05 $\pm$ 0.15, shown as solid dots in Fig.~5. This result was corroborated
by $\rho (T)$ measurements at constant fields, and the resulting jump
heights are given in Fig.~5 by open circles. At present it seems difficult
to give an unambiguous explanation for the doubling of $\rho$ upon the CJT
transition since several possibilities seem to apply equally well:\\

{\bf (i) Change of the double-exchange overlap integral}\\
It was argued \cite{Anane96} that the JT-elongation of MnO$_6$ octahedra and the exclusive
occupation of the $3z^2-r^2$ orbitals might lower the double-exchange
overlap integral along the elongated axis, while the overlap integrals
with neighbours along the perpendicular axes should become negligible.
This is somewhat doubtful, because JT distortions,
albeit fluctuating and in random directions, are present
already above T$_{JT}$. Furthermore, the
change in orthorhombicity at T$_{JT}$ might bring about a spontaneous
modification of the bond angle between Mn$^{3+}$, O$^{2-}$, and Mn$^{4+}$.
The importance of this angle for the absolute resistivity values (again
via the overlap integral) was pointed out in ref.~\cite{Fontcuberta}.
Neutron diffraction on undoped LaMnO$_3$ however has proven
that the tilt angle of MnO$_6$ octahedra (responsible for the buckling of
the Mn-O-Mn bonds) indeed increases with decreasing temperature - but
there is no discontinuous change around T$_{JT}$ \cite{Rodriguez}.\\

{\bf (ii) Decrease of the carrier localization volume}\\
Alternatively, we can consider a transport mechanism on the basis of
Mott's hopping concept, which is suggested by the low carrier mobility
resulting from a strong carrier localization \cite{Mott,Wagner}. The
resistivity in this model (and related hopping concepts) depends
exponentially on the ratio between the average hopping distance $R$ (at
high temperatures corresponding to the nearest neighbour distance) and the
half of the carrier localization length $L$. According to the orbital
structure sketched in Fig.~2 we might assume a localization of the hole-type
carrier within a volume $V_L$ extending from the Mn$^{4+}$ ion to the 6
nearest neighbours (via hybridization of orbitals) for the DJT state.
The CJT ordering (Fig.~3) will restrict $V_L$ to the Mn$^{4+}$ site and only
2 out of the 6 nearest neighbours. From this we can calculate for both
phases a localization length averaged over the three main crystallographic
directions, resulting in a resistive jump ratio of $\approx 1.70$. We point
out that this approach has to average over actually anisotropic localization
lengths, a situation for which the Mott-hopping mechanism is conceptually
not yet worked out.\\

{\bf (iii) Frustration of charge transport by orbital ordering}\\
The doubling of the resistivity may be understood by considering the
change in the allowed hopping-paths in the Figures 2 and 3. We postulate
that charge transport between Mn$^{3+}$ and Mn$^{4+}$ can only occur in
configurations with orbital overlap. This means that an $e_g$ electron
in a $x^2-y^2$ orbital can only be transferred to nearest-neighbour
Mn$^{4+}$ ions, which are located in the plane of this orbital, but not
in the perpendicular direction. An $e_g$ electron in a $3z^2-r^2$ orbital
can be transferred to nearest-neighbour Mn$^{4+}$ ions located on the axis
of this orbital, but not in the equatorial directions. We attribute to
both processes an equal, dimensionless conductivity $\sigma = 1$, and a
possible magnetic impedence of the charge-transfer by spin-misalignment
will preliminarily be ignored.\\

In the disordered state (Fig.~2) the hole associated with Mn$^{4+}$ can move
to all nearest Mn$^{3+}$ neighbours (due to low doping we consider only
Mn$^{3+}$), which provide orbital overlap, and is in principal mobile in
three dimensions. The probability for orbital overlap is, regardless of the
breathing mode and depending only on the disordered nature of the orbital
arrangement, given by the factor 1/2: 50 \%\ of the $e_g$-electrons occupy
$3z^2-r^2$ orbitals, which point with a 1/3 probability into the direction
of a given nearest-neighbour site. The other 50 \%\ of $e_g$ electrons
occupy $x^2-y^2$ orbitals with a respective statistical weight of 2/3. The
antiferrodistorsive order (Fig.~3) confines the charge carrier within the
$x^\star y^\star$ plane (movement perpendicular to this plane is forbidden
by the lack of suitable orbitals), and direct charge transfer within
this plane can occur only to 2 out of the 4 nearest neighbours.\\

Furthermore,
we need to take into account that the sample exhibits microtwinning in the
sense that the current path probes equal portions of differently oriented
domains. Therefore we calculate the relative conductivity averaged along
arbitrary directions of a three-dimensional cubic network, assuming that
domains with various orientations contribute to the total conductivity like
resistors connected in parallel. The average of conductivities can be
approximated by the average of the three types of directions shown in
Figure 6: (i) the principal axes of the crystal ($x^\star, y^\star$ and
$z^\star$ are used in the sense of Fig.~2); (ii) the square diagonals
($x^\star y^\star$, $x^\star z^\star$, and $y^\star z^\star$);
(iii) the cube diagonals.\\

In the three dimensional DJT state the conductivity of a Mn$^{4+}$ hole
along the principal axes is $\sigma^{3d}_p = 1/2$, i.e.~the elemental
conductivity $\sigma = 1$ (1 unit cell distance is spanned by 1 hopping
event) is weighted with the probability for a suitable orbital geometry.\\
\hspace*{5mm} The elemental conductivity along square diagonals is $1/\sqrt{2}$
(2 hops are required for a distance of $\sqrt{2}$ unit cells), and we take
into account that there are 2 possible hopping paths with a respective
probability for orbital overlap of $(1/2)^2$. The sqare-diagonal
conductivity is therefore $\sigma^{3d}_{sd} = 1/\sqrt{2} \cdot 2
\cdot (1/2)^2 \approx 0.354$.\\
\hspace*{5mm} Correspondingly, we obtain for cube diagonals $\sigma
= 1/\sqrt{3}$ (3 hops for a distance of $\sqrt{3}$ unit cells), and there
are now 6 possible paths with a probability factor of $(1/2)^3$. The
cube-diagonal conductivity is then $\sigma^{3d}_{cd} = 1/\sqrt{3} \cdot 6
\cdot (1/2)^3 \approx 0.433$.\\
\hspace*{5mm} The arithmetic average of $\sigma^{3d}_{p}, \sigma^{3d}_{sd}$,
and $\sigma^{3d}_{cd}$ approximates the averaged conductivity along all
possible directions and has a numerical value of
$\overline{\sigma^{3d}} \approx 0.429$.\\

The major difference in the more two-dimensional situation with
antiferrodistorsive order is the suppression of conductivity along the
$z^\star$ direction. Considering the conductivity along the two remaining
principal axes we note that, in the average, two steps are required to span
a distance of one unit cell in the desired direction, i.e.~the conductivity
is reduced by a factor of two. This is indicated by thin solid lines along
the black-shaded orbitals in Fig.~3. The corresponding average conductivity
is $\sigma^{2d}_p = 1/3 \cdot (0 + 1/2 + 1/2) = 1/3$. Due to the static
orbital arrangement we need not to consider the probability factors for
overlap geometries. The number of steps for charge transport along the diagonal
direction $x^\star y^\star$ is not altered by the orbital ordering (see
diagonal lines along the grey-shaded orbitals in Fig.~3), however the
conductivity along the other two square-diagonal directions becomes zero.
The average square-diagonal conductivity is therefore $\sigma^{2d}_{sd} =
1/3 \cdot(\sqrt{2}/2 + 0 + 0) \approx 0.236$. It is evident that the
cube-diagonal conductivity $\sigma^{2d}_{cd}$ is 0 and the total average
of all directions results in $\overline{\sigma^{2d}} \approx 0.190$.\\

The relative resistive jump height at the CJT transition is then given by
the ratio $\overline{\sigma^{3d}} / \overline{\sigma^{2d}} \approx 2.26$,
in a very close agreement with the experimentally found jump ratios between
1.9 and 2.2 (slightly dependent on the applied magnetic field).
We note that magnetoresistive measurements in untwinned single crystals,
with the current path along well-defined crystallographic axes, should bring
about jump ratios different of 2.\\

{\bf 3.3 Influence of the CJT transition on the paramagnetic susceptibility}\\

Associated with the doubling of the resistivity, we found also a sudden
drop in the paramagnetic magnetization, which is shown in the insert of
Fig.~1c. In field-dependent
magnetization measurements at constant temperatures (up to 5~T with a SQUID
magnetometer, and up to 50~T in pulsed magnetic fields, see Figs.~7 and 8)
this corresponds to a sudden, hysteretic upturn of the magnetization upon
leaving the CJT state, becoming unstable in sufficiently high fields. For
a better comparison, Fig.~7 includes also a magnetization curve at 272~K,
i.e.~above the JT ordering temperature, and a magnetization curve at 265~K,
i.e.~entirely inside the ordered phase. In addition to the presented
$M(B)$ data we studied magnetization curves at various temperatures between
220~K and 290~K. The signature of the CJT transition is especially
pronounced for the pulsed-fields magnetization technique (Fig.~8). As a
general tendency we can note that at lower temperatures (or respectively:
at higher magnetic fields) the hysteresis width shrinks rapidly and the
relative upturn of the magnetization decreases. Below 235~K and a
corresponding field of 18~T, we got no indication for a further
distinction between a JT-ordered and a disordered state.\\

To understand the influence of the CJT transition on the magnetic properties,
we should analyze the paramagnetic magnetization- and susceptibility data
not only in the framework of a simple model with independent magnetic ions,
but take also into account the existence of spin polarons.
The 'spin polaron' denotes hereby an ensemble of
neighbouring unit cells with parallel (on a local scale ferromagnetic) spin
alignment, which leads to a 'superparamagnetic behaviour'.
The existence of these clusters in CMR manganites was postulated
in ref.~\cite{Coey}, and found an experimental support in the articles
\cite{Wagner,DeTeresa,Chauvet}. In compounds with negligible orbital order,
i.e.~with higher doping concentration, the charge transport is governed
by the relative spin-misorientation between neighbouring spin polarons,
while the carriers are delocalized within their respective spin
cluster \cite{Wagner}.\\

For the size determination of spin polarons from the field-dependent
magnetization data (SQUID results in Fig.~7) we employed first a fitting
based on the Brillouin function $\cal{B}$, which is justified since the
respective temperatures around T$_{JT}$(B = 0) are sufficiently above the
ferromagnetic transition. The paramagnetic magnetization along the field
axis (per unit cell volume) is usually
given by  $M(B,T) = g \mu_B J \cdot {\cal B} \{ g \mu_B J B / k_B T \}$,
with the gyromagnetic ratio $g = 2$, $\mu_B$ the
Bohr magneton, and $J$ the spin-moment of the magnetic ions \cite{Ashcroft}.
For the mixture of Mn$^{3+}$ and Mn$^{4+}$ we replace $J$ by the
average $\overline{J} = 1.94$. In the case of superparamagnetic
clusters composed of n ions the $J$ factor in the argument
of the Brillouin function changes to $J^\star = n \cdot \overline{J}$
(causing a modification of the curvature) while the $\overline{J}$
in the prefactor remains unchanged. This is due to the compensation
of the $n$-fold increase in magnetic moment per (super-) paramagnetic
entity by the $1/n$-fold decrease of their density per unit volume.
Finally we have to replace the temperature $T$ by the effective temperature
scale $(T - T_C)^\gamma$, with T$_C$ = 188~K. Since we are not too
close to T$_C$ it is reasonable to chose $\gamma = 1$, according to
the Curie-Weiss law \cite{Ashcroft}.\\

The result of the fitting is given in Fig.~9 and suggests a CJT-induced
shrinking of superparamagnetic clusters from $J^\star = 8$ ($\approx$ 4 Mn
ions envolved) to $J^\star = 6$ (corresponding to roughly 3 Mn spins).
The three magnetization curves measured across the CJT phase boundary gave
hereby six data points. Besides this shrinking there is a slight tendency
towards an increasing cluster size with decreasing temperature. Describing
these cluster sizes on grounds of the orbital structures is difficult, since
a random orientation of $e_g$ electrons results, in principle, in
ferromagnetic nearest-neighbour correlations, while the antiferrodistorsive
structure promotes A-type antiferromagnetism \cite{Goodenough}. These
interactions are, however, weak because already in the case of pure
LaMnO$_3$, where CJT order is established around 800~K, the A-type AFM spin
order is only observed far below room temperature. We will therefore
restrict the discussion to the strongest ferromagnetic bonds, i.e.~the
coupling of the diluted Mn$^{4+}$ ions to the neighbouring Mn$^{3+}$ sites.
In the case of the ordered structure (see Fig.~3) the central Mn$^{4+}$
($J=3/2$) can undergo spin alignment in the sense of double exchange with
two out of the 6 Mn$^{3+}$ neighbours ($J=2$). The total $J^\star$ moment
of this entity corresponds to 5.5. For the DJT state (Fig.~2) 3 out of the
6 Mn$^{3+}$ neighbours are on the average in the $x^2-y^2$ configuration,
and the probability that an orbital lobe points towards the central Mn$^{4+}$
is 2/3. The other 3 orbitals are of the $3z^2-y^2$ shape, and the overlap
probability is 1/3 (according to the three possible spatial orientations
of these orbitals). The average total moment of this entity is therefore 7.5.
It might be accidental, but these figures (5.5 and 7.5) agree closely with
the experimentally found $J^\star$ values in the CJT and DJT phases.\\

As an alternative method for the determination of cluster sizes we analyzed
also the low-field magnetization data from Fig.~1c by means of the modified
(superparamagnetic) susceptibility formula $\chi = (g \mu_B)^2 \overline{J}
\cdot \{(J^\star + 1)/(3k_B (T - T_C))\}$ \cite{Ashcroft}.
The resulting $J^\star$ values (solid line in Fig.~9) are similar
to the afore-mentioned data, suggesting that the size of these clusters
is quite insensitive to the influence of external fields. From the
findings of this subsection we conclude that each hole-type charge-carrier,
associated with a Mn$^{4+}$ ion, is embedded into a locally ferromagnetic
environment, extending to the nearest-neighbour sites. The charge transfer
depends therefore indeed on the orbital configuration between Mn$^{4+}$ and
its nearest neighbours, while a possible impedence of the charge-movement
by magnetic disorder might, in a first approximation, be ignored.\\

{\bf 3.4 Competition between paramagnetic spin-alignment and
orbital ordering}\\

It can be seen from Fig.~10 that increasing magnetic fields shift the
cooperative JT transition, identified by the resistive jump and the
susceptibility drop, to lower temperatures. This shift is quadratic in
moderate fields (see insert of Fig.~10) with a slope of $- 0.11 K / T^2$,
in agreement with the behaviour published in ref.~\cite{Uhlenbruck}. It was
argued in earlier work on double-exchange that ferromagnetic spin alignment
enhances the mobility of charge carriers and lowers thereby their kinetic
energy in the sense of the uncertainty principle \cite{deGennes}. The
possible energy decrease due to itinerant behaviour can, certainly in a
ferromagnetic/quasimetallic state, exceed the total energy decrease
associated with a localized state forming a JT-deformed
environment \cite{Tyson}. Here we are dealing with a
similar problem.\\

In the higher conducting DJT state the
carriers are relatively mobile and charge transfer is impeded by spin
disorder. This spin-disorder persists in the CJT state, however the
static orbital structure results in an additional carrier localization,
as discussed in Sect.~3.2. The enhanced kinetic energy of the carrier
system is hereby overcompensated by the elastic energy of the lattice,
which is responsible for the transition to
the antiferrodistorsive structure. While external magnetic fields lower
the degree of spin-disorder localization in the CJT- and in the DJT state,
the gain in free energy is more pronounced in the latter case. The
frustration of carrier mobility in the CJT state would even persist in
the absence of spin disorder due to the static orbital arrangement.
In conclusion, magnetic fields lower the kinetic energy of the carrier
system more effectively in the DJT- than in the CJT state, and the DJT
state becomes stable in a wider temperature range, meaning that the
orbital ordering transition shifts to lower temperatures.\\

Possible measures of the field-induced decrease of free energy
in the DJT state are the enhancement of carrier mobility, and equivalently,
the decrease of resistivity. The CMR effect scales, in the paramagnetic state,
with the square of the Brillouin function, depending on the total moment of
superparamagnetic entities \cite{Wagner}, and the field-dependent
transition temperature T$_{JT}$(B) should scale according to the
implicit equation:

\begin{displaymath}
T_{JT}(B) = T_{JT}(B=0) - \alpha \cdot {\cal B}^2
\left(
\frac{g\mu_BJ^\star B}{k_B(T_{JT}(B) - T_C)}
\right)
\end{displaymath}

The $J^\star$ moment was chosen with 8 (experimental value right above
the phase transition in Fig.~9), and $\alpha$, the only free parameter,
connects the decrease of free energy (via delocalization) to the
CMR-induced lowering of the resistivity. The equation was solved
numerically for T$_{JT}$(B) and the best agreement with the data  (see
the solid line Fig.~10) was achieved for $\alpha$ = 47~K, with an uncertainty
of $\pm$ 2~K. The shape of the phase-boundary line is correctly
reproduced and the quadratic low-field behaviour emerges directly
from the properties of the Brillouin function. There are two alternative
mechanisms which can also explain a decrease of T$_{JT}$ with $B^2$,
but these effects are probably too small to account for the
observed shift:\\
\hspace*{5mm} Firstly, the intrinsic antiferromagnetism
of the antiferrodistorsive structure is caused by superexchange due to
the overlap of $t_{2g}$ orbitals along the $z^\star$ axis (Fig.~3).
External magnetic fields, or an internal Weiss field in the
ferromagnetic state, should therefore result in a decompression of
the $z^\star$ axis in order to minimize the superexchange-interaction,
which causes in turn a destabilization of the antiferrodistorsive order.
The correlations from superexchange are, however, weak and only notable
below T$_N$ $\approx$ 0.5 $\cdot$ T$_{JT}$.\\
\hspace*{5mm} Secondly, it is known that magnetic fields cause a shrinking
of the exponentially falling tails of electron wave functions
\cite{Shklovskii}. This shrinking reduces
the overlap between the $3z^2-r^2$ orbital and the two
opposite O$^{2-}$ orbitals and allows therefore for smaller JT-distortions,
with less pronounced minima in the free energy. The compression of the
Bohr radius $a$ is given by
$a \rightarrow a (1 - a^4 / (24 \lambda^4 + 3 a^4))$, with $\lambda$
being the 'Larmor length' $(\hbar / e B)^{1/2}$ \cite{Shklovskii}. This
expression is valid for $a << \lambda$ and predicts essentially
a shrinking proportional to the square of the local magnetic field.
Comparing the Bohr radius of $e_g$ electrons ($\approx$ 2 \AA ) with the
Larmor length at 50~T (36~\AA ) corresponds to a compression
effect of the order of 10$^{-7}$, which is insufficient
to affect the JT distortion itself.\\

It is noteworthy that the CJT transition vanishes for temperatures
below 235~K (see also Fig.~8), meaning that it might become second
order. The charge-ordering transition, characterized by the lifting of the
antiferrodistorsive structure and by ferromagnetic spin alignment,
shifts in non-zero fields to higher temperatures, because the internal
Weiss field is corroborated by the external contribution, see Fig.~10.
This means that sufficiently high fields can stabilize the nondistorsive
orbital structure (Fig.~4) already at temperatures above
T$_{co}$ = 147~K. We speculate that for temperatures where the
CJT transition apparently vanishes, see e.~g.~the pulsed-fields measurement
at 225~K in Fig.~8, the increasing external field can transform the
antiferrodistorsive order gradually to the nondistorsive type.
If there is a phase mixture between these two states, the typical
signatures of the CJT transition in magnetization and resistivity will
be smeared out and finally vanish, because transport- and magnetic
properties should be widely similar in the nondistorsive- and in the
disordered JT state.\\

{\bf 4.~Conclusions and Summary}\\
We investigated the resistive and magnetic behaviour of a rare-earth
manganite in the low-doping regime, La$_{7/8}$Sr$_{1/8}$MnO$_3$, where
the high Mn$^{3+}$ content results in a cooperative Jahn-Teller effect.
The phase transition from the orbitally disordered state to a phase with
antiferrodistorsive orbital order has a substantial influence on the
resistive as well as on the magnetic properties. The orbital ordering
impedes charge transfer along certain directions, resulting in a
doubling of the resistivity irrespective of external magnetic
fields and the corresponding spin-alignment and CMR effects.
The magnitude of this resistive jump was
calculated and found to be quantitatively correct on the basis of a
simple model, and crossing
the phase boundary from the orbitally ordered to the disordered state
corresponds to a strong, negative magnetoresistance effect,
which is independent of the CMR effect as such. The lower resistivity of
the disordered state goes along with an increase of the paramagnetic
susceptibility, which is especially remarkable, since it can only be
explained on grounds of superparamagnetic behaviour controlled
by ferromagnetic spin clusters or spin polarons. The typical cluster
moment was evaluated by two independent fitting procedures
and corresponds to the total moments of Mn$^{4+}$ ions and the surrounding
Mn$^{3+}$ sites, to which the Mn$^{4+}$ has an orbital overlap. The shift of
the orbital ordering transition to lower temperatures under the
influence of magnetic fields was explained by a stabilization of the
disordered phase through an enhancement of the carrier mobility.
Furthermore, we pointed out that the low-temperature properties of
La$_{7/8}$Sr$_{1/8}$MnO$_3$, which are incompatible with
antiferrodistorsive order, might be explained on grounds of a static
- but nondistorsive - orbital structure.\\[2ex]

{\bf Acknowledgements:} This work was supported by the Flemish Concerted
Action (GOA), the Fund for Scientific Research - Flanders (FWO), and the
Belgian Interuniversity Attraction Poles programs (IUAP). The authors thank
L.~Trappeniers for technical support with the pulsed-fields magnetization
measurements and V.~Bruyndoncx for help with the numerical solution of the
phase-boundary line. Constructive advice by R.~Gross, S.~Uhlenbruck, and
B.~B\"uchner from the University of Cologne is gratefully acknowledged.\\

\newpage

\begin{center}
{\Large \bf List of Figure Captions}\\[1ex]
\end{center}

{\bf Figure 1:} Temperature dependence of: (a) the lattice
constants of a La$_{7/8}$Sr$_{1/8}$MnO$_3$ single crystal (according to
ref.~\cite{Niemoeller}), (b) the resistivity, and (c)the low-field
magnetization. The cooperative JT transition at T$_{JT}$ = 269~K results in
a doubling of the resistivity and a drop of the paramagnetic susceptibility
(see insert in c). The gradual lifting of the cooperative JT distortion
below the Curie temperature (188~K) is associated with the step-like
development of a ferromagnetic/quasimetallic, and finally
ferromagnetic/charge-ordered state below T$_{CO}$ = 147~K. The dotted
line in (b) is a fit to the resistivity increase in the charge ordered
state by Shlovskii-Efros hopping.\\[3ex]

{\bf Figure 2:} The dynamic JT state is characterized by a random occupation
of $x^2-y^2$ and $3z^2-r^2$ orbitals and a vibronic mode (indicated by
arrows at the left $3z^2-r^2$ orbital), which favours an oscillation between
the two orbital types. The hole-type charge carrier associated with a
Mn$^{4+}$ ion is mobile in three dimensions. The orbitals in this and in
the subsequent Figures 3 and 4 are not drawn on scale with respect to the
lattice constant.\\[3ex]

{\bf Figure 3:} The antiferrodistorsive order lowers the conductivity along
the $x^\star$- and $y^\star$ axes by a factor of 2 (possible paths are
indicated by dark-shaded orbitals in part a), the conductivity along
the diagonal axes remains unaffected (grey-shaded orbitals). The
two-dimensional character prohibits charge transport along the $z^\star$
axis (part b) and the compression of the crystalline structure
along this axis causes A-type antiferromagnetism for undoped LaMnO$_3$
via superexchange between the $t_{2g}$ orbitals.\\[3ex]

{\bf Figure 4:} Tentative orbital structure of slightly doped LaMnO$_3$ at
low temperatures. The $x^2-y^2$ orbitals (50 \%\ occupation) are oriented
within the $x^\star z^\star$ plane, the $3z^2-r^2$ orbitals along the
$y^\star$ axis. The occupation of orbitals in subsequent layers along
$z^\star$ is reversed. The length $d$ corresponds to the mean diameter of
Mn$^{3+}$O$_6^{2-}$ octahedra without JT distortion. This nondistorsive
pattern is able to account for the reentrant structural properties
together with ferromagnetic spin alignment, in agreement with the
nearest-neighbour coupling rules \cite{Goodenough}.\\[3ex]

{\bf Figure 5:} Normalized magnetoresistance (left axis) at the transition
from the cooperative JT state in low fields to the dynamic JT state in
higher fields. Upon lowering temperature the width of the hysteresis loops
decreases rapidly, indicating a weakening of the first-order character of
the CJT-transition under the presence of external magnetic fields. The
relative height of the resistive jump from the ordered- to the disordered
state (right axis) is almost field- and temperature independent with an
absolute value around 2.1. Solid dots refer to the magnetoresistance
measurements at constant temperature, open dots to temperature sweeps at
fixed magnetic fields. The dotted line is a guide to the eye.\\[2ex]

\newpage

{\bf Figure 6:} Illustration of nearest-neighbour hopping on a cubic
network with different conductivities voor principal axes, square-, and
cube diagonals. The elemental conductivities ($\sigma = 1, 1/\sqrt{2},
1/\sqrt{3}$) have to be weighted by the number of possible paths, and by
the probability for a suitable orbital-overlap configuration.\\[3ex]

{\bf Figure 7:} SQUID magnetization measurements around the CJT-transition:
the orbitally ordered state becomes unstable at higher fields, resulting in
a hysteretic upturn of the paramagnetic magnetization. For clarity, the
absolute values of the magnetization curves at different temperatures
are shifted by the indicated offset values.\\[3ex]

{\bf Figure 8:} The pulsed-fields magnetization measurement at 245~K
is qualitatively equivalent to the curves in Fig.~7. The signature
of the phase transition vanishes abruptly for temperatures lower than
235~K (here shown for 225~K). The orbital-ordering transition results
in pronounced spikes in the non-integrated $dM/dB$ signal
(see insert).\\[3ex]

{\bf Figure 9:} The temperature-dependent size of preformed ferromagnetic
spin clusters in the paramagnetic phase around the CJT transition. The data
points were determined by Brillouin fits to the magnetization curves (Fig.~7)
and suggest a shrinking from 4 to 3 Mn ions involved in the formation of an
individual spin cluster. Measurements crossing the phase-boundary line gave
two corresponding data points (solid squares and circles). The solid line
was calculated on the basis of the low-field magnetization data from
Fig.~1c.\\[3ex]

{\bf Figure 10:} Increasing magnetic fields shift the CJT transition
to lower temperatures (quadratic in low fields, compare insert) and the
charge ordering transition to higher temperatures (the dotted line is a
guide to the eye). The orbital ordering line is fitted on grounds of
the field-induced mobility contribution to the free energy of the
DJT state. The distinction between orbitally ordered and disordered state
ceases above 20~T. The squares correspond to resistive data from $\rho(T)$
measurements at constant fields, the open circles to pulsed-field
magnetization.\\


\newpage

\begin{thebibliography}{99}

\bibitem{Kusters} R.~M.~Kusters, J.~Singleton, D.~A.~Keen, R.~McGreevy,
and W.~Hayes, {\sl Physica B} {\bf 155}, 362 (1989), and R.~von Helmolt,
J.~Wecker, B.~Holtzapfel, L.~Schultz, and K.~Samwer,
{\sl Phys.~Rev.~Lett.} {\bf 71}, 2331 (1993).

\bibitem{Ramirez} A.~P.~Ramirez, {\sl J.~Phys.~Condens.~Matter}, {\bf 9},
8171 (1997).

\bibitem{Imada} M.~Imada, A.~Fujimori, and Y.~Tokura, {\sl Rev.~Mod.~Phys.}
{\bf 70}, 1039 (1998).

\bibitem{deGennes} P.~G.~de Gennes, {\sl Phys.~Rev.} {\bf 118}, 141 (1960)
and references therein.

\bibitem{Millis} A.~J.~Millis, {\sl Phys.~Rev.~B} {\bf 53}, 8434 (1996) and
{\sl Nature} {\bf 392}, 147 (1998).

\bibitem{Mueller} E.~M\"uller-Hartmann and E.~Dagotto,
{\sl Phys.~Rev.~B} {\bf 54}, R6819 (1996).

\bibitem{Wagner} P.~Wagner, I.~Gordon, L.~Trappeniers, J.~Vanacken, F.~Herlach,
V.~V.~Moshchalkov, and Y.~Bruynseraede, {\sl Phys.~Rev.~Lett.} {\bf 81},
3980 (1998).

\bibitem{Radaelli} P.G.~Radaelli, D.~E.~Cox, M.~Marezio, and S.-W.~Cheong,\\
{\sl Phys.~Rev.~B} {\bf 55}, 3015 (1997).

\bibitem{Care} for a review on Wigner crystallization see:
C.~M.~Care and N.~H.~March,\\ {\sl Adv.~Phys.} {\bf 24}, 101 (1975).

\bibitem{Kuwahara} H.~Kuwahara, Y.~Tomioka, A.~Asamitsu, Y.~Moritomo, and
Y.~Tokura,\\ {\sl Science} {\bf 270}, 961 (1995).

\bibitem{Tomioka} Y.~Tomioka, A.~Asamitsu, Y.~Moritomo, H.~Kuwahara,
and Y.~Tokura, {\sl Phys.~Rev.~Lett.} {\bf 74}, 5108 (1995).

\bibitem{Kawano97} H.~Kawano, R.~Kajimoto, H.~Yoshizawa, Y.~Tomioka,
H.~Kuwahara, and Y.~Tokura, {\sl Phys.~Rev.~Lett.} {\bf 78}, 4253 (1997).

\bibitem{Pinsard} L.~Pinsard, J.~Rodriguez-Carvajal, A.~H.~Moudden, A.~Anane,
A.~Revcolevschi, and C.~Dupas, {\sl Physica B} {\bf 234 - 236}, 856 (1997).

\bibitem{Uhlenbruck} S.~Uhlenbruck, R.~Teipen, R.~Klingeler, B.~B\"uchner,
O.~Friedt, M.~H\"ucker, H.~Kierspel, T.~Niem\"oller, L.~Pinsard,
A.~Revcolevschi, and R.~Gross, {\sl Phys.~Rev.~Lett.} {\bf 82}, 185 (1999).

\bibitem{Niemoeller} T.~Niem\"oller, M.~v.~Zimmermann, J.~R.~Schneider,
S.~Uhlenbruck, B.~B\"uchner, T.~Frello, N.~H.~Andersen, L.~Pinsard,
A.~M.~de L\'eon-Guevara, and A.~Revcolevschi,
{\sl Eur.~Phys.~J.~B} {\bf 8}, 5 (1999).

\bibitem{Gehring} G.~A.~Gehring and K.~A.~Gehring, {\sl Rep.~Prog.~Phys.}
{\bf 38}, 1 (1995).

\bibitem{Goodenough} J.~B.~Goodenough, A.~Wold, R.~J.~Arnott, and N.~Menyuk,
{\sl Phys.~Rev.} {\bf 124}, 373 (1961) and
J.~B.~Goodenough and J.~M.~Longo, {\sl Landolt-B\"ornstein, Neue Serie
III/4a}, 126 (Springer Verlag, Berlin, 1970).

\bibitem{Murakami} Y.~Murakami, J.~P.~Hill, D.~Gibbs, M.~Blume,
I.~Koyama, M.~Tanaka, H.~Kawata, T.~Arima, Y.~Tokura, K.~Hirota,
and Y.~Endoh, {\sl Phys.~Rev.~Lett.} {\bf 81}, 582 (1998).

\bibitem{Rodriguez} J.~Rodriguez-Carvajal, M.~Hennion, F.~Moussa, L.~Pinsard,
and A.~Revcolevschi, {\sl Physica B} {\bf 234 - 236}, 848 (1997).

\bibitem{Anane95} A.~Anane, C.~Dupas, K.~Le Dang, J.~P.~Renard, P.~Veillet,
A.~M.~de L\'eon-Guevara, F.~Millot, L.~Pinsard, and A.~Revcolevschi,
{\sl J.~Phys.: Condens.~Matter} {\bf 7}, 7015 (1995), and references
therein.

\bibitem{Herlach} F.~Herlach, C.~C.~Agosta,R.~Bogaerts, W.~Boon, I.~Deckers,
A.~De Keyzer, N.~Harrison, A.~Lagutin, L.~Li, L.~Trappeniers, J.~Vanacken,
L.~Van Bockstal, and A.~Van Esch, {\sl Physica B} {\bf 216}, 161 (1996).

\bibitem{Kawano96} H.~Kawano, R.~Kajimoto, M.~Kobota, and H.~Yoshizawa,\\
{\sl Phys.~Rev.~B} {\bf 53}, R14 709, (1996).

\bibitem{Neumeier} J.~J.~Neumeier, K.~Andres, and K.~J.~McClellan,
{\sl Phys.~Rev.~B} {\bf 59}, 1701 (1999).

\bibitem{Anane96} A.~Anane, C.~Dupas, K.~Le Dang, J.~P.~Renard, P.~Veillet,
L.~Pinsard, and A.~Revcolevschi, {\sl Appl.~Phys.~Lett.} {\bf 69},
1160 (1996).

\bibitem{Moreo} A.~Moreo, S.~Yunoki, and E.~Dagotto,
{\sl Science} {\bf 283}, 2034 (1999).

\bibitem{Efros} A.~L.~Efros and B.~I.~Shklovskii, {\sl J.~Phys.~C, Solid
State Physics}, {\bf 8}, 249 (1975).

\bibitem{Aliev} F.~G.~Aliev, E.~Kunnen, K.~Temst, K.~Mae, G.~Verbanck,
J.~Barnas, V.~V.~Moshchalkov, and Y.~Bruynseraede,
{\sl Phys.~Rev.~Lett.} {\bf 78}, 134 (1997) and references therein.

\bibitem{Tyson} T.~A.~Tyson, J.~Mustre de Leon, S.~D.~Conradson,
A.~R.~Bishop, J.~J.~Neumeier, H.~R\"oder, and Jun Zang,
{\sl Phys.~Rev.~B} {\bf 53}, 13985, (1996).

\bibitem{Maezono} R.~Maezono, S.~Ishihara, and N.~Nagaosa,
{\sl Phys.~Rev.~B} {\bf 57}, R13993 (1998).

\bibitem{Fontcuberta} J.~Fontcuberta, B.~Mart\'{\i}nez, A.~Seffar,
S.~Pi\~{n}ol, J.~L.~Garc\'{\i}a-Mu\~{n}oz, and X.~Obradors,
{\sl Phys.~Rev.~Lett.} {\bf 76}, 1122 (1996).

\bibitem{Mott} N.~F.~Mott and E.~A Davis, {\sl Electronic Processes in
Non-Crystalline Materials} (Clarendon Press, Oxford, 1979).

\bibitem{Coey} J.~M.~D.~Coey, M.~Viret, L.~Ranno, and K.~Ounadjela,\\
{\sl Phys.~Rev.~Lett.} {\bf 75}, 3910 (1995).

\bibitem{DeTeresa} J.~M.~De Teresa, M.~R.~Ibarra, P.~A.~Algarabel,
C.~Ritter, C.~Marquina, J.~Blasco, J.~Garcia, A.~Delmoral, and
Z.~Arnold, {\sl Nature} {\bf 386}, 256 (1997).

\bibitem{Chauvet} O.~Chauvet, G.~Goglio, P.~Molinie, B.~Corraze,
and L.~Brohan,\\ {\sl Phys.~Rev.~Lett.} {\bf 81}, 1102 (1998).

\bibitem{Ashcroft} N.~W.~Ashcroft and N.~D.~Mermin, {\sl Solid State Physics},
(Saunders College, Philadelphia, 1976).

\bibitem{Shklovskii} B.~I.~Shklovskii and A.~L.~Efros, {\sl Electronic
Properties of Doped Semiconductors}, Springer Series in Solid-State
Sciences {\bf 45} (1984).

\end{thebibliography}
\end{document}